\documentclass[12pt]{article}

\newsavebox{\sboxpubnumber}
\newsavebox{\sboxpubdate}
\newcommand{\pubdate}[1]{\begin{lrbox}{\sboxpubdate}{#1}\end{lrbox}}

\newcommand{\Title}[1]{\begin{center} {\Large #1 } \end{center}}
\newcommand{\Author}[1]{\begin{center}{ \sc #1} \end{center}}
\newcommand{\Address}[1]{\begin{center}{ \it #1} \end{center}}

\newenvironment{Abstract}{\begin{quotation}  }{\end{quotation}}
\newenvironment{Presented}{\begin{quotation} \begin{center}
             PRESENTED AT\end{center}\bigskip
      \begin{center}\begin{large}}{\end{large}\end{center}
      \end{quotation}}
\newcommand{\Acknowledgements}{\bigskip  \bigskip \begin{center} \begin{large}
             \bf ACKNOWLEDGEMENTS \end{large}\end{center}}
\newcommand{\simlt}
{\mbox{\raisebox{-0.5ex}{$\textstyle \; \sim$}
\raisebox{ 0.8ex}{$\textstyle  \!\!\!\!\!\!\! <$  }}}
\newcommand{\simgt}
{\mbox{\raisebox{-0.5ex}{$\textstyle \; \sim$}
\raisebox{ 0.8ex}{$\textstyle  \!\!\!\!\!\!\! >$  }}}
\begin{document}

\begin{titlepage}
\pubdate{\today}                    

\vfill
\Title{
Does Dark Matter
at the Center and in the Halo of the Galaxy
Consist of the Same Particles?
}
\vfill
\Author{Neven Bili\'{c}\footnote{Permanent address:
Rudjer Bo\v skovi\'c Institute,
P.O. Box 180, 10002 Zagreb, Croatia \\
\hspace*{5mm} Email: bilic@thphys.irb.hr}\ ,
Faustin Munyaneza, Gary B.\ Tupper, \\ and Raoul D.\
Viollier\footnote{Email: viollier@physci.uct.ac.za}
}
\Address{
Institute of Theoretical Physics and Astrophysics \\
 Department of Physics, University of Cape Town  \\
 Private Bag, Rondebosch 7701, South Africa \\
 }
\vfill
\begin{Abstract}
After a discussion of the properties of degenerate fermion balls,
we analyze the orbits of the star S0-1,
which has the smallest projected
distance to Sgr A$^{*}$,
in the supermassive black hole as well as in the
fermion ball scenarios of the Galactic center.
It is shown that both scenarios are consistent with the data,
as measured during the last six years by
Genzel {\it et al.} and Ghez {\it et al.}.
We then consider a self-gravitating ideal
fermion gas at nonzero temperature as a model for the
Galactic halo.
The Galactic halo of mass
$\sim 2 \times 10^{12} M_{\odot}$ enclosed within
a radius of $\sim 200$
kpc implies the existence of a
supermassive compact dark object at
the Galactic center that is in hydrostatic and
thermal equilibrium with the halo.
The central object has a maximal mass of $\sim 2.3 \times 10^{6}
M_{\odot}$ within a minimal radius of $\sim 18$ mpc
or $\sim$ 21 light-days for
fermion masses $\sim 15$ keV.
We thus conclude that both the
supermassive compact dark object
and the halo could be made of the
same weakly interacting $\sim 15$ keV particle.
\end{Abstract}
\vfill
\begin{Presented}
    COSMO-01 \\
    Rovaniemi, Finland, \\
    August 29 -- September 4, 2001
\end{Presented}
\vfill
\end{titlepage}
\def\thefootnote{\fnsymbol{footnote}}
\setcounter{footnote}{0}

\section{Introduction}
In the past,
self-gravitating degenerate
neutrino matter has been suggested as a model
for quasars,
with neutrino
masses in the $0.2 {\rm keV} \simlt m \simlt 0.5  {\rm MeV}$ range
\cite{1}.
Later it was used to describe dark matter in clusters of galaxies and
in galactic halos,
with neutrino masses in the
$ 1 \simlt m/{\rm eV} \simlt 25  $ range
\cite{cos}.
More recently,
supermassive compact
dark objects consisting of weakly interacting degenerate
fermionic matter, with fermion masses
in the $10 \simlt m/{\rm keV}\simlt 20$
range, have been proposed
\cite{2,3,4,5,6}
 as an alternative to the
supermassive black holes that are
believed to reside at the centers of many
galaxies.
It has been pointed out that such degenerate fermion balls
could cover
\cite{4}
 the whole range of the supermassive compact dark
objects that have been observed so far
with masses ranging
from $10^6$ to $3 \times 10^9\, M_{\odot}$
\cite{kor}.
Most recently, it has been
shown that a weakly
interacting dark matter particle in the mass
range $1 \simlt m/{\rm keV} \simlt 5$
could solve the problem of the excessive
structure generated on subgalactic scales in
$N$-body and hydrodynamical
simulations of structure formation in this Universe \cite{bode}.

So far the masses of $\sim 20$ supermassive
compact dark objects at the center
of galaxies have been measured
using various techniques \cite{kor}.
The most massive compact dark object ever observed is
located at the center of M87 in the Virgo cluster,
and it has a mass of about
$3 \times 10^9 M_{\odot}$ \cite{7}.
If we identify this object
of maximal mass with a degenerate fermion ball
at the Oppenheimer-Volkoff (OV) limit \cite{opp},
i.e., $M_{\rm{OV}} = 0.54 M_{\rm Pl}^3\,
m^{-2} g^{-1/2} \simeq 3 \times 10^9
M_{\odot}$
\cite{4},
where $M_{\rm{Pl}}=\sqrt{\hbar c/G}$
denotes the Planck mass,
 this allows us to fix the fermion mass to $m \simeq 15$ keV
for a spin and particle-antiparticle degeneracy factor of $g=2$.
Such a relativistic object
would have a radius of $R_{\rm{OV}}= 4.45
R_{\rm{S}} \simeq$ 1.5 light-days
(1.19 light-days = 1 mpc),
where $R_{\rm{S}}$ is the Schwarzschild
radius of the mass $M_{\rm{OV}}$.
It would thus be virtually indistinguishable from
a black hole of the same mass,
as the closest stable orbit around a black hole
has a radius of 3 $R_{\rm{S}}$ anyway.

Near the lower end of the observed mass  range
is the compact dark object
located at the Galactic center
\cite{eck}
 with a mass of $M_{\rm c} \simeq 2.6 \times
10^6 M_{\odot}$.
Interpreting this object as a degenerate fermion ball consisting of
$m \simeq 15$ keV and $g= 2$ fermions,
the radius is $R \simeq 21$ light-days
$\simeq 7 \times 10^4 R_{\rm S}$
\cite{2},
$R_{\rm S}$ being the Schwarzschild
radius of the mass $M_{\rm c}$.
Such a nonrelativistic object is far from being a black hole.
The shallowness of
its gravitational potential may be probed by infalling gas
and stars orbiting within this fermion ball.
A particle at infinity at rest falling
towards the fermion ball can merely  acquire
the escape velocity
of $\sim 1700$ km/s
from the center of the
fermion ball.
Therefore the spectrum of the radiation
emitted by the infalling gas will
be cut off at  high frequencies, as is actually seen
in the far infrared
spectrum of the
enigmatic radio source Sgr A$^{*}$ at the
Galactic center
\cite{10}.
From the observed cut-off at $\sim 10^{13}$ Hz,
corresponding to emission at
$\sim$ 10 light-days,
in a thin accretion disk scenario,
one can deduce
a lower limit for the radius of
$R \simgt$ 18 light-days
\cite{5}.
Similarly, the observed motion of stars within a projected distance of
$\sim$ 6 to $\sim$ 50 light-days from Sgr A$^{*}$
\cite{eck}
yields, apart from the
mass, an upper limit for
the radius of the fermion ball $R \simlt$ 22
light-days.

The required weakly
interacting fermion of $\sim$ 15 keV mass cannot be an
active neutrino,
as it would overclose the Universe by orders of magnitude
\cite{11}.
Moreover, an active neutrino of
$\sim$ 15 keV is  disfavored by
the experimental
data on solar and atmospheric neutrinos, as these are
most probably
oscillating into
active neutrinos with small $\delta m^2$ \cite{fuk}, and
the $\nu_{\rm e}$ mass has been determined to be $< 3$ eV
\cite{rev}.
However, the $\:\sim\! 15$ keV fermion could very well be a sterile
neutrino, contributing
$\Omega_{\rm d} \simeq$ 0.3 to the dark matter
fraction of the critical density today.
Indeed, as has been shown for an
initial lepton asymmetry of
$\sim 10^{-3}$,  a sterile neutrino of mass
$\sim$ 10 keV may be
resonantly produced in the early Universe with near
closure density, i.e. $\Omega_{\rm d} \sim$ 1
\cite{shi}.
The resulting energy spectrum
of the sterile neutrinos is
cut off for energies larger than the resonance
energy, thus mimicking
a degenerate fermion gas.
As an alternative possibility,
the  $\sim$ 15 keV sterile neutrino
could be replaced by the axino
\cite{cov}
or the gravitino
\cite{din,mur}
in soft supersymmetry breaking scenarios.

In the recent past,
galactic halos have been successfully modeled as
a self-gravitating
isothermal gas of particles of arbitrary mass,
the density of which  scales asymptotically as
$r^{-2}$, yielding flat rotation curves
\cite{col}.
As the supermassive compact
dark objects at the galactic centers
are well described by a gas of fermions
of mass $m \sim 15$ keV at $T = 0$,
it is tempting to explore the
possibility that one could describe
both the supermassive compact dark objects
and their galactic halos in
a unified way in terms of a fermion gas
at finite temperature.
We will show in this paper that this
is indeed the case, and that
the observed dark matter distribution in the
Galactic halo is consistent with
the existence of a supermassive compact dark
object at the center of
the Galaxy which has about the right mass and size,
and is in thermal and hydrostatic equilibrium
with the halo.

\section{Dynamics of the Stars Near the Galactic Center}
We now would like to compare the predictions of the black hole and fermion
ball
scenarios of the
Galactic center, for the stars with
the smallest projected distances to Sgr A$^{*}$,
based on the measurements of their positions
during the last six years \cite{6,eck}.
The projected orbits of three stars, S0-1 (S1),
S0-2 (S2) and S0-4 (S4), show deviations from
uniform motion on a straight line during the
last six years, and they thus may contain
nontrivial information about the potential.
For our analysis we have selected the star,
S0-1, because its projected distance from
Sgr A$^{*}$ in 1995.53, 4.4 mpc or 5.3 light-days,
makes it most likely that it could be orbiting
within a fermion ball of radius $\sim$ 18 mpc
or $\sim$21 light-days. We thus may in principle
distinguish between the black hole and
fermion ball scenarios for this star.

The dynamics of the stars in the gravitational
field of the supermassive compact dark object
can be studied solving Newton's equation of
motion, taking into account the initial position
and velocity vectors at, e.g., $t_{0}$ = 1995.4 yr,
i.e., $\vec{r} (t_{0}) \equiv (x,y,z)$ and $\dot{\vec{r}} (t_{0})
\equiv (v_{x},v_{y},v_{z})$. For the fermion ball
the source of gravitational field is
the mass ${\cal{M}}(r)$ enclosed within a radius $r$ \cite{2,6}
while  for the black hole it is $M_{c} = {\cal{M}}(R_{c})
= 2.6\times 10^6 M_{\odot}$.
The $x$-axis is chosen in the direction opposite to the right
ascension (RA), the $y$-axis in the direction
of the declination, and the $z$-axis points towards the sun.
The black hole and the center of the fermion ball
are assumed to be at the position of Sgr A$^{*}$ which
is also the origin of the coordinate system at an assumed
distance of 8 kpc from the sun.

In
Figs.\ \ref{fig1} and \ref{fig2}  the right ascension
(RA) and declination of S0-1
are plotted as a function of time
for various unobservable
$z$'s and $v_z=0$ in 1995.4,
 for the black hole and fermion ball
scenarios.
The velocity components
$v_{x}$ = 340 km s$^{-1}$ and $v_{y} = -1190$ km s$^{-1}$
 in 1995.4
 have been fixed from observations.
In the case of a black hole, both RA and declination depend
strongly on  $z$ in 1995.4, while the $z$-dependence
of these quantities
in the fermion ball scenario is rather weak.
We conclude that the RA and declination data of S0-1 are well fitted with
$|z| \approx 0.25''$ in the black hole scenario, and with
$|z| \simlt 0.1''$ in the fermion ball case
($1'' =38.8$ mpc = 46.2 light-days at 8 kpc).
Of course, we can also
try to fit the data varying both the unknown radial velocity
$v_{z}$  and the unobservable radial distance $z$.
The results are
summarized in Fig.\ \ref{fig3}, where  the $z - v_{z}$ phase-space of
1995.4, that fits the data, is shown.
The small range of acceptable $|z|$ and $|v_{z}|$
values in the black hole scenario (solid vertical line) reflects the fact
that the orbit of S0-1 depend strongly on  $z$.
The weak sensitivity  of the orbit on $z$ in the fermion ball
case is the reason for the much larger $z - v_{z}$ phase-space fitting
the  data of S0-1 \cite{eck},
as shown by the dashed box. The dashed and
solid curves describe the just bound orbits in the fermion ball and black hole
scenarios, respectively.
The star
S0-1 is unlikely to be unbound, because in the absence of close
encounters with stars of the central cluster,
S0-1 would have to fall in with an initial
velocity that is inconsistent with the velocity dispersion of the stars at
infinity.

Fig.\ \ref{fig4} shows some typical projected orbits of S0-1 in the black hole and
fermion ball scenarios. The data of S0-1 may be fitted in both scenarios with
appropriate choices of $v_{x}$, $v_{y}$, $z$ and $v_{z}$ in 1995.4. The
inclination angles of the orbit's plane, $\theta$ = arccos $\left(
L_{z}/|\vec{L}| \right)$, with $\vec{L} = m \vec{r} \times \dot{\vec{r}}$,
are shown next to the orbits. The minimal inclination angle that describes
the data in the black hole case is $\theta$ = 70$^{o}$, while in the fermion
ball scenario it is $\theta$ = 0$^{o}$. In the black hole case, the minimal
and maximal distances from Sgr A$^{*}$ are $r_{\rm min}$ = 0.25$''$ and
$r_{\rm max}$ = 0.77$''$, respectively, for the orbit with $z= 0.25''$ and
$v_{z}$ = 0 which has a period of $T_0 \approx$ 161 yr. The orbits with $z$ =
0.25$''$ and $v_{z}$ = 400 km s$^{-1}$ or $z$ = 0.25$''$ and $v_{z}$ = 700 km
s$^{-1}$ have periods of $T_0 \approx$ 268 yr or $T_0 \approx$ 3291 yr,
respectively. In the fermion ball scenario, the open orbit with $z = 0.1''$
and $v_{z}$ = 0 has a ``period'' of $T_0 \approx$ 77 yr with $r_{\rm min}=
0.13''$ and  $r_{\rm max}= 0.56''$. The open orbits with $z = 0.1''$ and
$v_{z}$ = 400 km s$^{-1}$ or $z = 0.1''$ and $v_{z}$ = 900 km s$^{-1}$ have
``periods'' of $T_0 \approx$ 100 yr or $T_0 \approx$ 1436 yr,
respectively.

In concluding, it is important to note that, based on the data
of the star S0-1
\cite{eck},  the fermion ball scenario cannot be ruled out.
In fact, in view of the $z - v_{z}$ phase space, that is much larger in
the fermion ball scenario than in the black hole case, there is reason to
treat the fermion ball scenario of the supermassive compact dark object at the
center of our Galaxy with the respect it deserves.

\section{Dark Matter in the Center and the
Halo of the Galaxy}
Degenerate fermion balls are well understood
in terms of the Thomas-Fermi theory applied
to self-gravitating fermionic matter at $T = 0$
\cite{2}.
Extending this theory to
nonzero temperature
\cite{16,bil,17},
it has been shown
that at some critical
temperature $T = T_{\rm c}$,
a self-gravitating ideal fermion gas, having a mass
below the  OV limit  enclosed in a
spherical cavity of radius $R$, may
undergo a first-order gravitational
phase transition from a diffuse state to a
condensed state.
This is best seen plotting the energy  and free
energy as functions of
the temperature which are three-valued in some
temperature interval,
exhibiting a Maxwell-Boltzmann branch at high
temperatures and the degenerate branch at low temperatures.
However, this first-order
phase transition can only take place
if the Fermi gas is able to get rid
of the large latent heat which is
due to the binding energy of the fermion ball.
As the short-range interactions
of the fermions are negligible, the gas cannot
release its latent heat;
it will thus be trapped for temperatures $T <
T_{\rm c}$ in a thermodynamic
quasi-stable supercooled state
close to the point of gravothermal
collapse.
The Fermi gas will be caught in the supercooled
state even if the total mass of the gas exceeds the  OV limit,
as a
stable condensed state does not exist in this case.

The formation of a supercooled state
close to the point of
gravothermal collapse,
may be understood as a process similar
to that of violent relaxation,
which was introduced to describe rapid
virialization of stars of different mass in globular clusters
\cite{lynd,bin}
without invoking binary collisions of the stars,
as these would not contribute
significantly to thermalization on a
scale of the age of the Universe.
Through the gravitational collapse of a cold
overdense fluctuation, $\sim$ 1 Gyr after the Big Bang,
part of gravitational energy transforms into the kinetic energy
of random motion of small-scale density fluctuations.
The resulting virialized
cloud will thus be well
approximated by a gravitationally stable thermalized
halo.
In order to
estimate the mass-temperature ratio,
we assume that the cold overdense cloud of the
mass of the Galaxy $M$
 stops expanding at the time $t_{\rm m}$,
reaching its maximal radius
$R_{\rm m}$ and minimal
average density $\rho_{\rm{m}}= 3 M/(4 \pi R_{\rm{m}}^3)$.
 The total energy per particle is just the gravitational energy
 \begin{equation}
 E=-\frac{3}{5}\frac{GM}{R_{\rm{m}}} \, .
 \label{eq001}
 \end{equation}
 Assuming spherical collapse
 \cite{padma}
 one arrives at
 \begin{equation}
 \rho_{\rm{m}}=\frac{9\pi^2}{16} \bar{\rho}(t_{\rm{m}})
 =\frac{9\pi^2}{16} \Omega_{\rm{d}} \rho_0 (1+z_{\rm{m}})^3,
 \label{eq002}
 \end{equation}
 where $\bar{\rho}(t_{\rm{m}})$ is the background density
 at the time $t_{\rm{m}}$ or cosmological
 redshift $z_{\rm{m}}$, and
$\rho_0\equiv 3 H_0^2/(8\pi G)$ is the present critical
 density.
 We now
 approximate the virialized cloud by
a singular isothermal sphere
\cite{bin}
of mass $M$ and
radius $R$,
characterized by
a constant circular velocity
$ \Theta=(2 T/m)^{1/2}$
and the density profile
$ \rho(r)=\Theta^2/4\pi G r^2 .$
Its total energy per particle is the sum of gravitational
and thermal energies, i.e.,
 \begin{equation}
 E=-\frac{1}{4}\frac{GM}{R}
 =-\frac{1}{4}\Theta^2 .
 \label{eq004}
 \end{equation}
 Combining Eqs. (\ref{eq001}), (\ref{eq002}),
 and (\ref{eq004}),
 we find
 \begin{equation}
 \Theta^2=\frac{6\pi}{5} G
 (6 \Omega_{\rm{d}} \rho_0 M^2)^{1/3}(1+z_{\rm{m}}) .
 \label{eq005}
 \end{equation}
 Taking $\Omega_{\rm{d}}=0.3$, $M=2\times 10^{12} M_{\odot}$,
 $z_{\rm{m}}=4$, and $H_0=65\,{\rm km\, s^{-1} Mpc^{-1}}$, we find
 $\Theta \simeq 220\, {\rm km\, s^{-1}}$, which corresponds to the
 mass-temperature ratio $m/T\simeq 4\times 10^6$.

Next, we briefly discuss the general-relativistic
extension of the Tho\-mas-Fer\-mi theory
\cite{bil}
for a self-gravitating gas
of $N$ fermions with mass $m$
and degeneracy factor $g$
at the temperature $T$
enclosed
in  a sphere of radius $R$.
We denote
by $p$, $\rho$, and $n$
the  pressure,
energy density, and particle number density
of the gas,
respectively.
In the following we use the units in which
$G=1$.
The metric generated by the mass distribution
is  static, spherically symmetric, and asymptotically
flat, i.e.,
\begin{equation}
ds^2=\xi^2 dt^2 -(1-2{\cal{M}}/r)^{-1} dr^2 -
     r^2(d\theta^2+\sin \theta d\phi^2).
\label{eq13}
\end{equation}
For numerical convenience,
we introduce the parameter
$\alpha=\mu/T$
and the substitution
$\xi=(\varphi+1)^{-1/2} \mu/m$,
where $\mu$ is the chemical potential associated with the
conserved particle number $N$.
The equation
of state for a self-gravitating gas may thus be represented
in parametric form \cite{ehl} as
\begin{equation}
n=\frac{1}{\pi^2}
               \int^{\infty}_{0} dy
\frac{y^2}
{1+\exp \{[(y^2+1)^{1/2}/(\varphi+1)^{1/2}-1]\alpha\} },
\label{eq83}
\end{equation}
\begin{equation}
\rho=\frac{1}{\pi^2}
               \int^{\infty}_{0}dy\,
\frac{y^2(y^2+1)^{1/2}}
{1+\exp \{[(y^2+1)^{1/2}/(\varphi+1)^{1/2}-1]\alpha\} },
\label{eq84}
\end{equation}
\begin{equation}
p=\frac{1}{3\pi^2}
               \int^{\infty}_{0}dy\,\frac{y^4
(y^2+1)^{-1/2}}
{1+\exp \{[(y^2+1)^{1/2}/(\varphi+1)^{1/2}-1]\alpha\} },
\label{eq85}
\end{equation}
where appropriate length and mass  scales
$a$ and $b$,  respectively, have been chosen such that
$ a=b=
(2/g)^{1/2}/m^2$.
Restoring $\hbar$, $c$, and $G$,
we have
\begin{equation}
a=
\sqrt{\frac{2}{g}} \,
\frac{\hbar M_{\rm{Pl}}}{c m^2}
=1.0798 \times 10^{10}\,
\sqrt{\frac{2}{g}} \,
\left(
\frac{15{\rm keV}}{m}
\right)^2
{\rm km},
\end{equation}
\begin{equation}
b=
\sqrt{\frac{2}{g}} \,
\frac{M_{\rm{Pl}}^3}{m^2}
=0.7251 \times 10^{10}\,
\sqrt{\frac{2}{g}} \,
\left(
\frac{15{\rm keV}}{m}
\right)^2
M_{\odot} \, .
\end{equation}
Thus fermion mass, degeneracy factor,  and
chemical potential are eliminated from the
equation of state.

Einstein's field equations for the metric (\ref{eq13})
are given by
\begin{equation}
\frac{d\varphi}{dr} =
-2(\varphi+1)\frac{{\cal M}+4\pi r^3 p}{r(r-2{\cal{M}})} \, ,
\label{eq88}
\end{equation}
\begin{equation}
\label{eq89}
\frac{d{\cal{M}}}{dr}=4\pi r^2 \rho.
\end{equation}
To these two equations we add
\begin{equation}
\frac{d{\cal N}}{dr}=4\pi r^2 (1-2{\cal{M}}/r)^{-1/2} n
\label{eq93}
\end{equation}
imposing
 particle-number conservation as
a condition at the boundary
\begin{equation}
{\cal N}(R)=N.
\label{eq94}
\end{equation}
Eqs. (\ref{eq88})-(\ref{eq93})
should be integrated using
the bo\-un\-da\-ry conditions at the origin, i.e.,
\begin{equation}
\varphi(0)=\varphi_0 > -1
\, , \;\;\;\;\;
{\cal{M}}(0)=0
\, , \;\;\;\;\;
{\cal{N}}(0)=0.
\label{eq90}
\end{equation}

It is useful to introduce the degeneracy
parameter
$\eta=\alpha \varphi/2$,
which,
in the Newtonian limit,
approaches
$\eta_{\rm nr}=(\mu_{\rm nr} -V)/T$,
with $\mu_{\rm nr}=\mu-m$ being
the nonrelativistic chemical potential
and $V$ the Newtonian potential.
As $\varphi$ is monotonously decreasing with increasing
 $r$, the strongest degeneracy is obtained at the center
 with $\eta_0=\alpha\varphi_0/2$.
The parameter $\eta_0$,
uniquely related to the central
density and pressure,
will eventually be fixed
by the requirement (\ref{eq94}).
For $r\geq R$, the function $\varphi$ yields
the usual empty-space Schwarzschild
solution
\begin{equation}
\varphi(r)=\frac{\mu^2}{m^2}
\left(1-\frac{2 M}{r}\right)^{-1}-1\, ,
\label{eq91}
\end{equation}
  with
\begin{equation}
M={\cal M}(R)=\int_0^R dr 4\pi r^2 \rho(r) .
\label{eq92}
\end{equation}
Given the temperature $T$, the set of self-consistency
equations (\ref{eq83})-(\ref{eq93}), with the boundary
conditions (\ref{eq94})-(\ref{eq92})
defines the general-relativistic extension of the Thomas-Fermi
equation.

\section{Numerical Results}
The numerical procedure is now straightforward.
For a fixed, arbitrarily chosen
$\alpha$,
 we first  integrate
Eqs.
 (\ref{eq88}) and (\ref{eq89})
 numerically
on the interval $[0,R]$ to find
the solutions
 for various central values
 of the degeneracy parameter
$\eta_0$.
Integrating (\ref{eq93}) simultaneously,
yields
${\cal N}(R)$ as a function of $\eta_0$.
We then select the
value of $\eta_0$
for which
${\cal N}(R)=N$.
The chemical potential $\mu$
corresponding to
this particular solution
 is given by
Eq. (\ref{eq91})
which in turn yields
the
parametric dependence on the temperature
through $\alpha=\mu/T$.

The quantities
$N$, $T$,and $R$  are free parameters of our model
and their
range of values are dictated
by the physics of the problem at hand.
At $T=0$
the number of fermions $N$  is
 restricted by the OV limit
$N_{\rm OV}=2.89
\times 10^{9}\,
\sqrt{2/g}
(15\,{\rm keV}/m)^2
M_{\odot}/m $.
However, at nonzero temperature, stable solutions exist
with $N>N_{\rm OV}$, depending on temperature
and radius.
In the following  $N$ is required to be
of the order $2\times 10^{12} M_{\odot}/m$,
so that for any $m$, the total mass
is close to the estimated mass of the halo
\cite{wilk}.
As we have demonstrated,
the expected particle mass-temperature ratio of the halo
is given by $\alpha\simeq m/T=4\times 10^4$.
 The halo radius
$R$ is in principle unlimited; in practice, however,
it should not exceed half the average intergalactic distance.
It is known that an
 isothermal configuration has no natural boundary,
in contrast to the degenerate
 case of zero temperature,  where
for given $N$ (up to the OV limit)
the radius $R$ is naturally fixed by the condition of
vanishing pressure and density.
At nonzero  temperature,
with $R$ being unbounded, our gas would occupy
the entire space, and
fixing $N$ would make
$p$ and $\rho$
vanish everywhere.
Conversely,
if we do not fix $N$ and integrate the
equations  on the interval $[0,\infty)$,
both $M$ and $N$ will
diverge at infinity for $T>0$.
Thus, one is forced to introduce a cutoff.
In an isothermal model of a similar kind
\cite{cha},
the  cutoff was set
at the radius $R$, where the energy density
was by about six orders of magnitude smaller than the central value.
Our choice of
$R=200$ kpc
is based on the estimated size of the Galactic halo.

The only remaining free parameters of our model are the fermion
mass $m$ and the degeneracy factor $g$,
which always appear in the combination
$m^4g$.
We fix these parameters at
$m=15$ keV and $g=2$, and justify this choice {\em a posteriori}.

The results of numerical integration of Eqs.
 (\ref{eq88}) and (\ref{eq89}),
 without restricting $N$, are presented
in Fig.\ \ref{fig5}, where we plot the  particle number $N$
as a function of the central
degeneracy parameter
$\eta_0$ for several
values of $\alpha$ close to $4\times 10^6$.
For fixed $N$, there is a range of $\alpha$,
where the Thomas-Fermi equation has multiple solutions.
For example, for $N=2\times 10^{12}$ and
$\alpha=4\times 10^6$ six solutions are found,
which are  denoted by
(1), (2), (3), (3'), (2'), and (1')
corresponding to the values $\eta_0 =$
 -30.528,
 -25.354,
 -22.390,
  29.284,
  33.380, and
  40.479, respectively.
In Figs. \ref{fig6}
and \ref{fig7} we plot the corresponding density profiles and enclosed
masses, respectively.
For negative central value $\eta_0$,
for which the degeneracy parameter is negative everywhere,
the system behaves basically as a
Maxwell-Boltzmann isothermal sphere.
Positive values of the central degeneracy parameter $\eta_0$
are characterized by a pronounced central core
of mass of about $2.5 \times 10^6 M_{\odot}$
within a radius of about 20 mpc.
The presence of this core is obviously due to
the degeneracy pressure of the
Fermi-Dirac statistics.
A similar structure was obtained in
collisionless stellar systems modeled as
a nonrelativistic Fermi gas
\cite{chav}.

Figs. \ref{fig6}
and \ref{fig7} show two important features.
First,
a galactic halo at a given temperature $T$
may or may not have a central core
depending whether  the central degeneracy parameter $\eta_0$
is positive or negative.
Second,
the closer to zero $\eta_0$ is,
the smaller the radius is at which the
$r^{-2}$  asymptotic behavior of the density begins.
The flattening of the Galactic rotation  curve
begins in the range  $1 \simlt r/{\rm kpc} \simlt 10$,
hence the solution (3') most likely describes the
Galaxy's halo.
This may be verified by calculating the rotational
curves in our model.
We know already from the estimate (\ref{eq005})
that our model
yields the correct asymptotic circular velocity of
220 km/s.
In order to make a more realistic comparison
with the observed Galactic rotation curve,
we must include
two additional matter components: the bulge and
the disk.
The bulge is modeled as a spherically symmetric matter distribution
of the form
\cite{you}
\begin{equation}
\rho_{\rm b}(s)=\frac{e^{-hs}}{2s^3}
\int_0^{\infty} du
\frac{e^{-hsu}}{[(u+1)^8-1]^{1/2}} \, ,
\label{eq006}
\end{equation}
where $s=(r/r_0)^1/4$, $r_0$ is the effective radius of the bulge
and $h$ is  a parameter.
We adopt  $r_0=2.67$ kpc
and $h$ yielding the bulge mass
$M_{\rm b}= 1.5 \times 10^{10} M_{\odot}$
\cite{suc}.
In Fig. \ref{fig8}  the mass
of halo and bulge enclosed within
a given radius is plotted for various $\eta_0$.
Here, the gravitational backreaction  of the bulge on
the fermionic halo has been taken into account.
The data points, indicated  by squares, are
the  mass
$M_{\rm c}=2.6 \times 10^6 M_{\odot}$ within
18 mpc, estimated from the motion of the stars
near Sgr A$^*$ \cite{eck},
and the mass
$M_{50}=5.4^{+0.2}_{-3.6}\times 10^{11}$
within 50 kpc,
 estimated from
the motions
of satellite galaxies and globular clusters
\cite{wilk}.
Variation of the central degeneracy parameter
$\eta_0$ between 24 and 32 does not change
the essential halo features.

In Fig.\ \ref{fig9} we plot
the circular velocity components of
the halo, the bulge, and the disk.
The contribution of the disk
 is modeled  as \cite{per}
\begin{equation}
\Theta_{\rm d}(r)^2=
\Theta_{\rm d}(r_{\rm o})^2
\frac{1.97 (r/r_{\rm o})^{1.22}}{
[(r/r_{\rm o})^2+0.78^2]^{1.43}} \, ,
\label{eq007}
\end{equation}
where we take
$r_{\rm o}=13.5$ kpc and
$\Theta_{\rm d}=100$ km/s.
Here it is assumed for simplicity that
the disk does not influence the mass distribution
of the bulge and the halo.
Choosing the central degeneracy
$\eta_0=28$ for the halo, the data
by Merrifield and Olling \cite{oll} are reasonably well
fitted.

We now turn to the discussion of
our choice of the fermion mass $m=15$ keV
for a degeneracy factor $g=2$.
To this end
we investigate how  the mass of the
central object,
i.e., the mass $M_{\rm c}$ within 18 mpc,
depends on $m$ in the interval
5 to 25 keV,
for various
$\eta_0$.
We find
that $m\simeq 15$ keV gives always the maximal value of
$M_{\rm c}$
ranging between 1.7 to 2.3 $\times 10^6 M_{\odot}$
for $\eta_0$ between 20 and 28.
Hence, with  $m\simeq 15$ keV  we get the
value closest to the mass of the central object
$M_{\rm c}$
estimated from the motion of the stars
near Sgr A$^*$ \cite{eck}.

We  now present the results of the calculations for  fixed
particle number and  temperatures near the
point of gravothermal collapse.
In Fig.\ \ref{fig10} the energy per particle
defined as $E=M/N-m$
is plotted
as a function of temperature for fixed
$N=2\times 10^{12}$.
The plot looks very much like that of
a canonical Maxwell-Boltzmann  ensemble \cite{bin},
with one important difference:
in the Maxwell-Boltzmann case the curve  continues to
spiral inwards {\em ad infinitum} approaching the point
of the singular isothermal sphere,
that is
characterized by an infinite central
density.
In Fermi-Dirac case the spiral consists of two  almost
identical curves.
The inwards winding of the spiral begins
for some  negative central degeneracy and stops at the point
$T=2.3923 \times 10^{-7} m$,
$E=-1.1964 \times 10^{-7} m$,
where $\eta_0$ becomes zero.
This part of the curve, which basically depicts the behavior
of a nondegenerate gas, we call
{\em Maxwell-Boltzmann branch}.
By increasing the central degeneracy parameter
further to positive values,
the spiral begins to unwind outwards very close to
the inwards winding curve.
The outwards winding curve
will eventually depart from the
Maxwell-Boltzmann branch for temperatures $T \simgt 10^{-3} m$.
Further increase of the central degeneracy parameter brings us to
a region, where general-relativistic effects become
important.
The curve will exhibit another spiral
for temperatures and energies of the order of a few $10^{-3}m$
approaching the limiting temperature  $T_{\infty}= 2.4 \times 10^{-3}m$
and energy
$E_{\infty}= 3.6 \times 10^{-3}m$ with both  the central degeneracy
parameter
and the central density approaching infinite values.
It is remarkable that
gravitationally stable configurations
with  arbitrary large central degeneracy parameters
exist
at finite temperature
even though the total
mass exceeds the OV limit by several
orders of magnitude.

\section{Conclusions}
In summary,
using the Thomas-Fermi
theory, we have shown that
a  weakly interacting
fermionic gas  at finite temperature
yields  a mass distribution that
successfully describes both the center and the halo
of the Galaxy.
For a fermion mass
$m \simeq 15$ keV,
a reasonable fit to the rotation
curve is achieved with the
temperature $T = 3.75$ meV and
the degeneracy parameter
at the center $\eta_0=28$.
With the same parameters,
we obtain
the mass
$M_{50} = 5.04\times 10^{11} M_{\odot}$
and
$M_{200} = 2.04\times 10^{12} M_{\odot}$
within 50 and 200 kpc, respectively.
These values agree quite well with the mass estimates
based on the motions
of satellite galaxies and globular clusters
\cite{wilk}.
Moreover, the mass
of $M_{\rm c} \simeq 2.27\times 10^6 M_{\odot}$,
enclosed within 18 mpc,
agrees reasonably  well
with the observations of the compact dark object at
the center of the Galaxy.
We thus conclude that both the Galactic halo and center
could be made of the same fermions.

An observational consequence of this unified scenario of
fermion ball and fermion halo at
finite temperature could be the
direct observation of the radiative decay of
the fermion (assumed here to be a sterile neutrino)
into a standard neutrino,
i.e., $f \rightarrow \nu \gamma$.
The X-ray luminosity of the compact dark
object is most easily  observed.
If the lifetime for the decay
$f \rightarrow \nu \gamma$ is 0.82 $\times
10^{19}$ yr,
the luminosity of a $M_{\rm c} = 2.6 \times 10^6 M_{\odot}$
fermion ball
would be 0.9 $\times 10^{34}$ erg s$^{-1}$.
This is
consistent with the upper limit
of the X-ray luminosity of
$\sim$ (0.5 - 0.9) $\times 10^{34}$ erg s$^{-1}$
of the source with radius $0.5''$ $\simeq$ 23 light-days,
whose center nearly
coincides with Sgr A$^{*}$,
as seen by the Chandra satellite in the
2 to 7 keV band \cite{bag}.
The lifetime is proportional to
$\sin^{-2} \theta$, $\theta$ being the
unknown mixing angle of the sterile
with active neutrinos.
With a lifetime of 0.82 $\times$ 10$^{19}$ yr we obtain
an acceptable value for the mixing
angle squared of $\theta^2 = 1.4
\times 10^{-11}$.
The X-rays originating from such a radiative decay would
contribute about two orders of magnitude
less than the observed diffuse X-ray
background at this wavelength if the
sterile neutrino is the dark matter
particle of the Universe.
The signal
observed at the Galactic center
would be a sharp X-ray line at $\sim$ 7.5
keV for $g = 2$
and  $\sim$ 6.3
keV for $g = 4$.
This line could be misinterpreted as the  Fe $K_{\alpha}$ line at
6.67 keV.
Scattering with baryonic
matter within the Galactic center could distribute
the energy more evenly in the 2 to 7 keV band.
The X-ray luminosity
would be tracing the fermion matter distribution, and it
could thus be an important test of the fermion ball scenario.
Of course the angular
resolution would need to be $\simlt 0.1''$
and the sensitivity would have to
extend beyond 7 keV.

\Acknowledgements

This
research is in part supported by the Foundation of Fundamental
Research (FFR) grant number PHY99-01241 and the Research Committee of
the University of Cape Town.  The work of N.B. is supported in part by
the Ministry of Science and Technology of the Republic of Croatia
under Contract No. 00980102.

\newpage
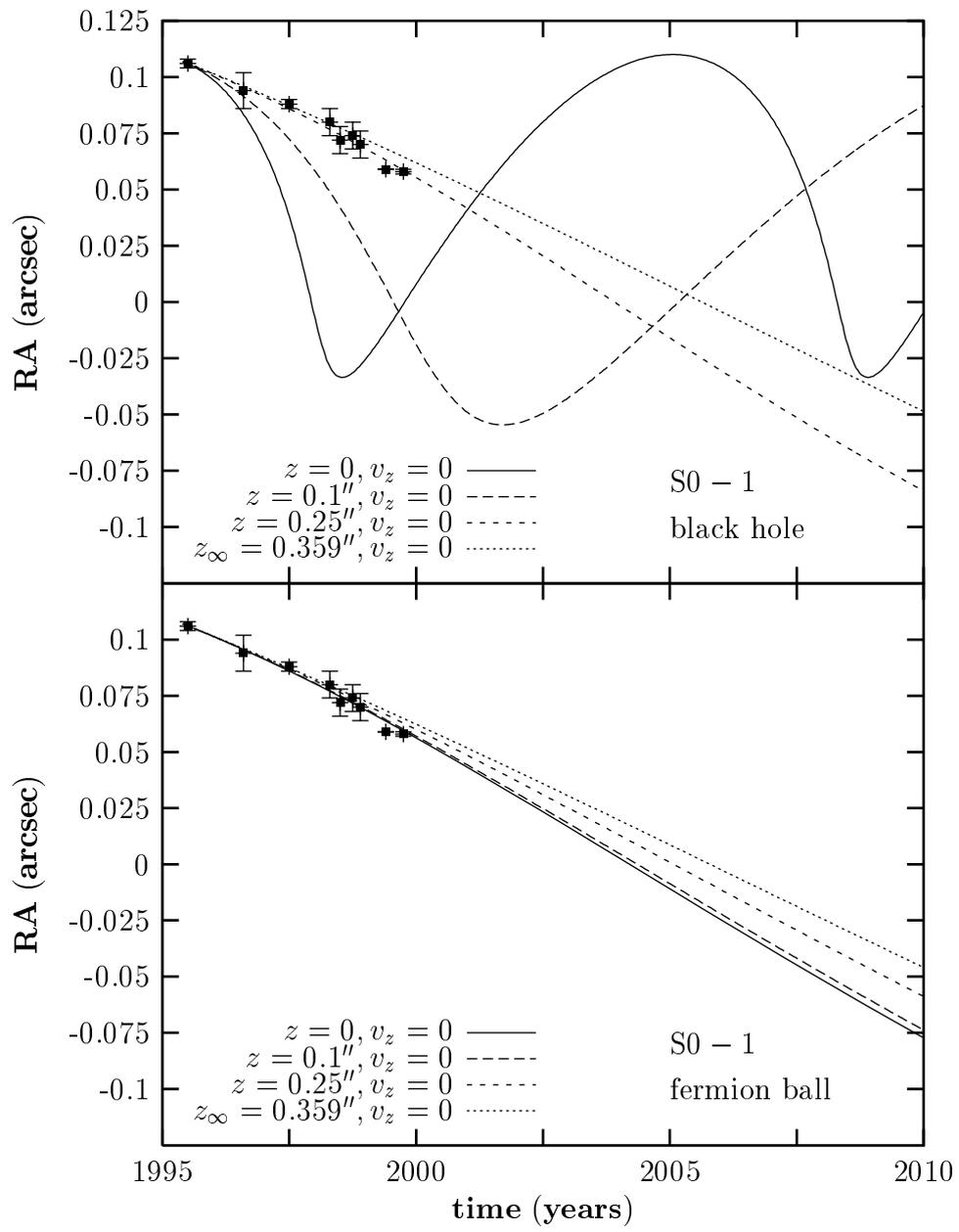
\begin{figure}
\caption{
Right ascension of S0-1 versus time for
various $|z|$ and
$v_{x}$ = 340 km s$^{-1}$, $v_{y} = -1190$ km s$^{-1}$
and
$v_{z}$ = 0
in 1995.4.
}
\label{fig1}
\end{figure}
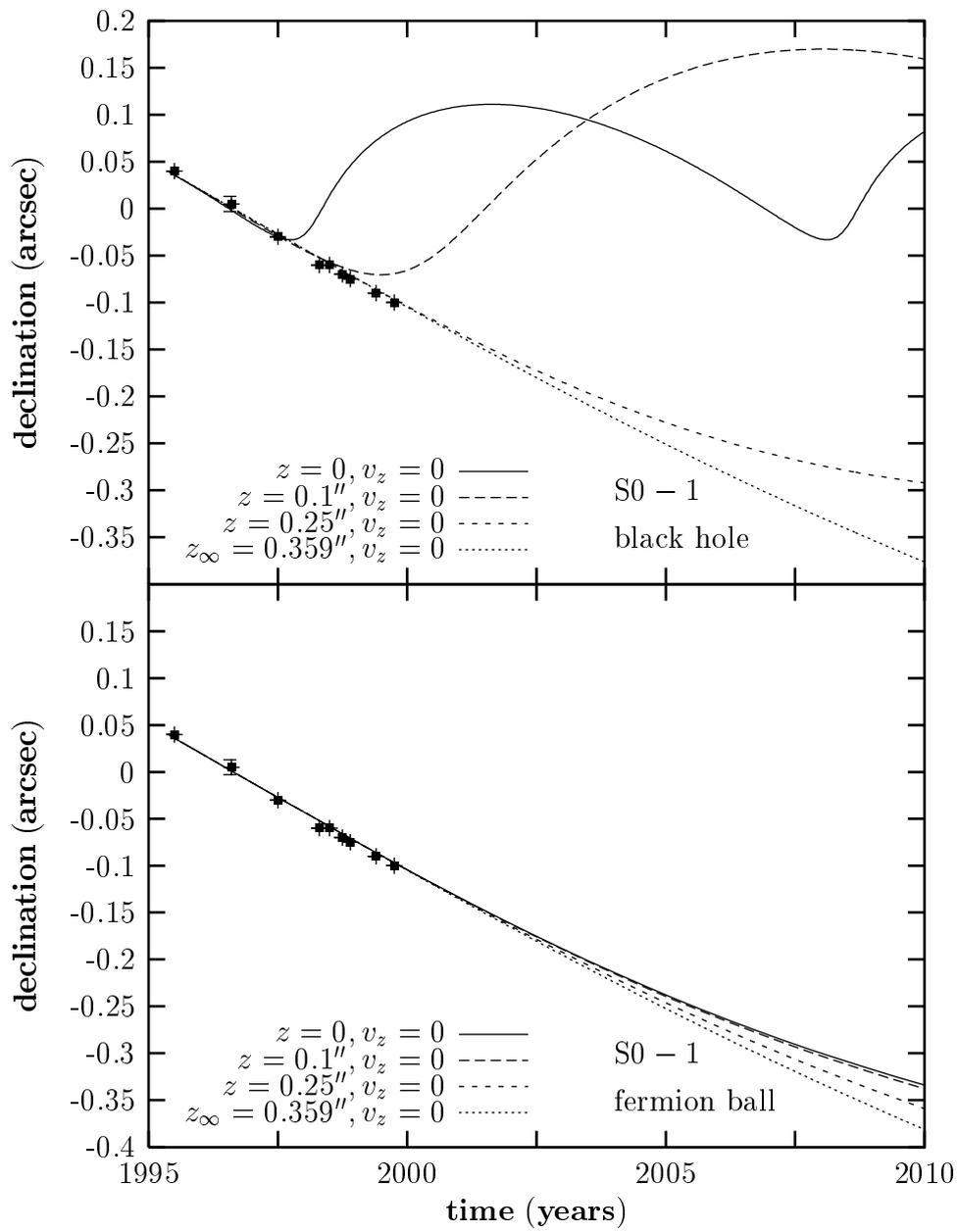
\begin{figure}
\caption{
Declination of S0-1
versus time for
various $|z|$ and
$v_{x}$ = 340 km s$^{-1}$, $v_{y} = -1190$ km s$^{-1}$
and
$v_{z}$ = 0
in 1995.4.
}
\label{fig2}
\end{figure}
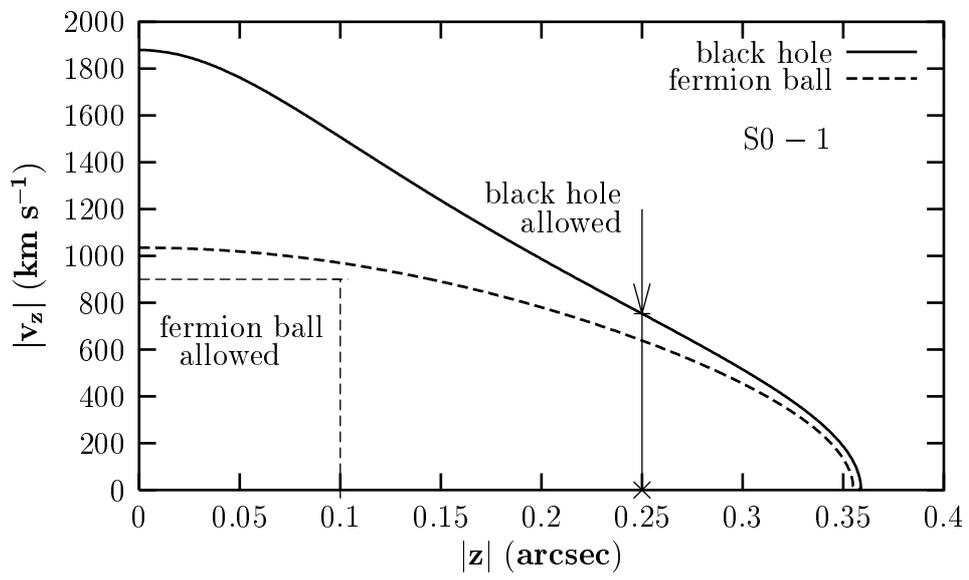
\begin{figure}
\caption{
The $z - v_{z}$ phase-space
 that fits the S0-1 data.
}
\label{fig3}
\end{figure}
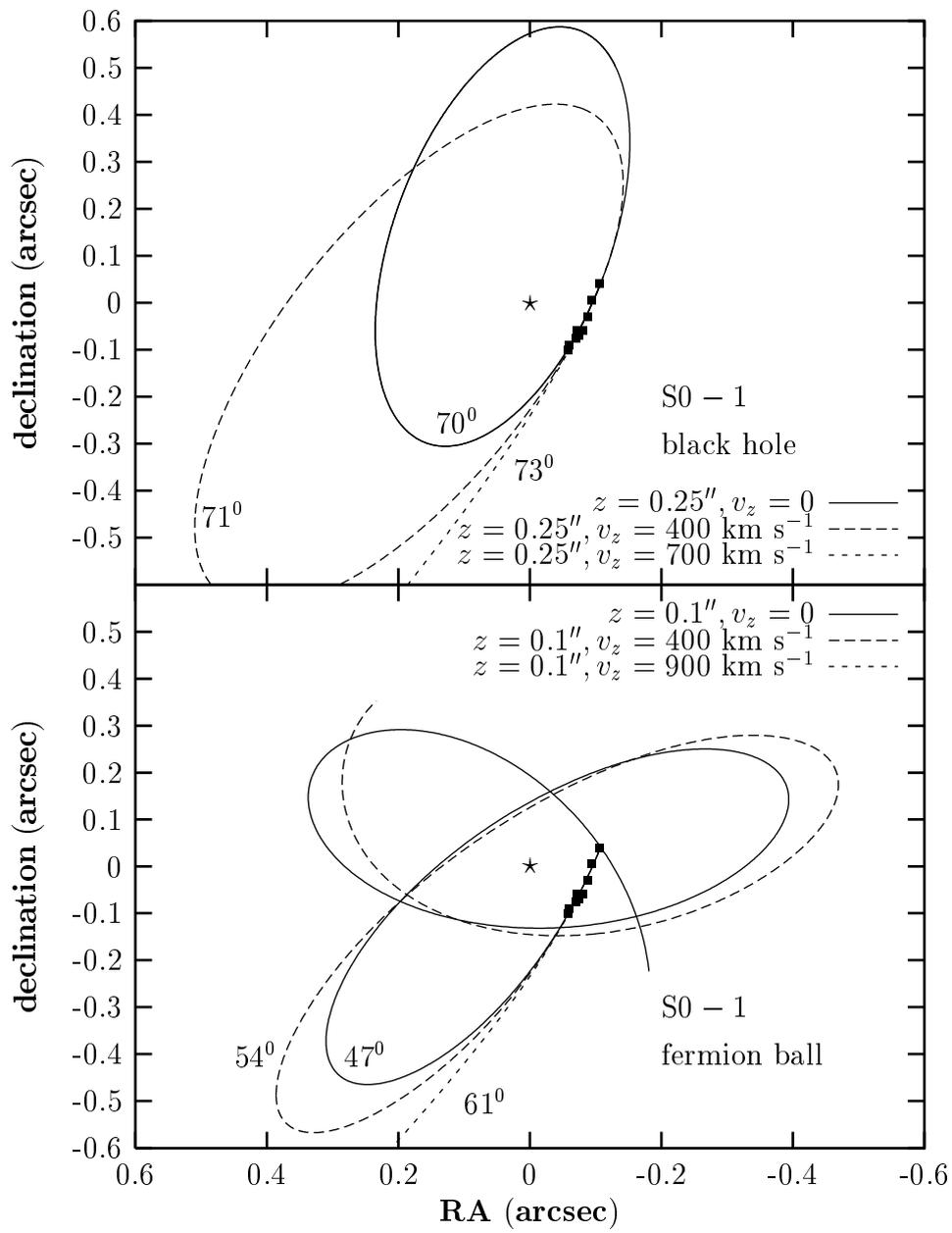
\begin{figure}
\caption{
Examples of typical orbits of S0-1.
}
\label{fig4}
\end{figure}
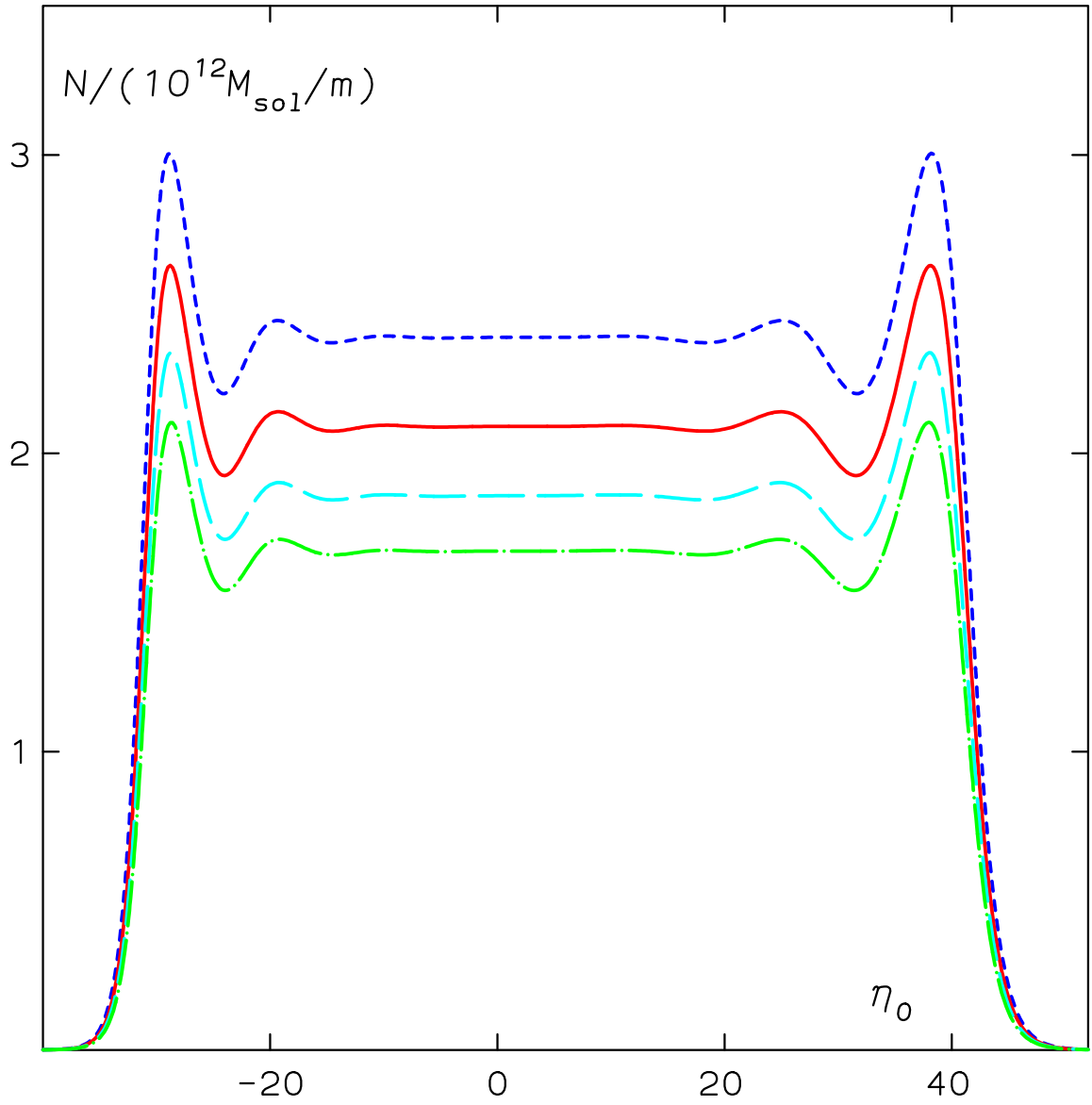
\begin{figure}
\caption{
Number of particles  versus central degeneracy
parameter
for $m/T= 4\times 10^6$
(solid),
$3.5\times 10^6$
(short dashs),
$4.5\times 10^6$
(long dashs), and
$5\times 10^6$
 (dot-dashed line).
}
\label{fig5}
\end{figure}
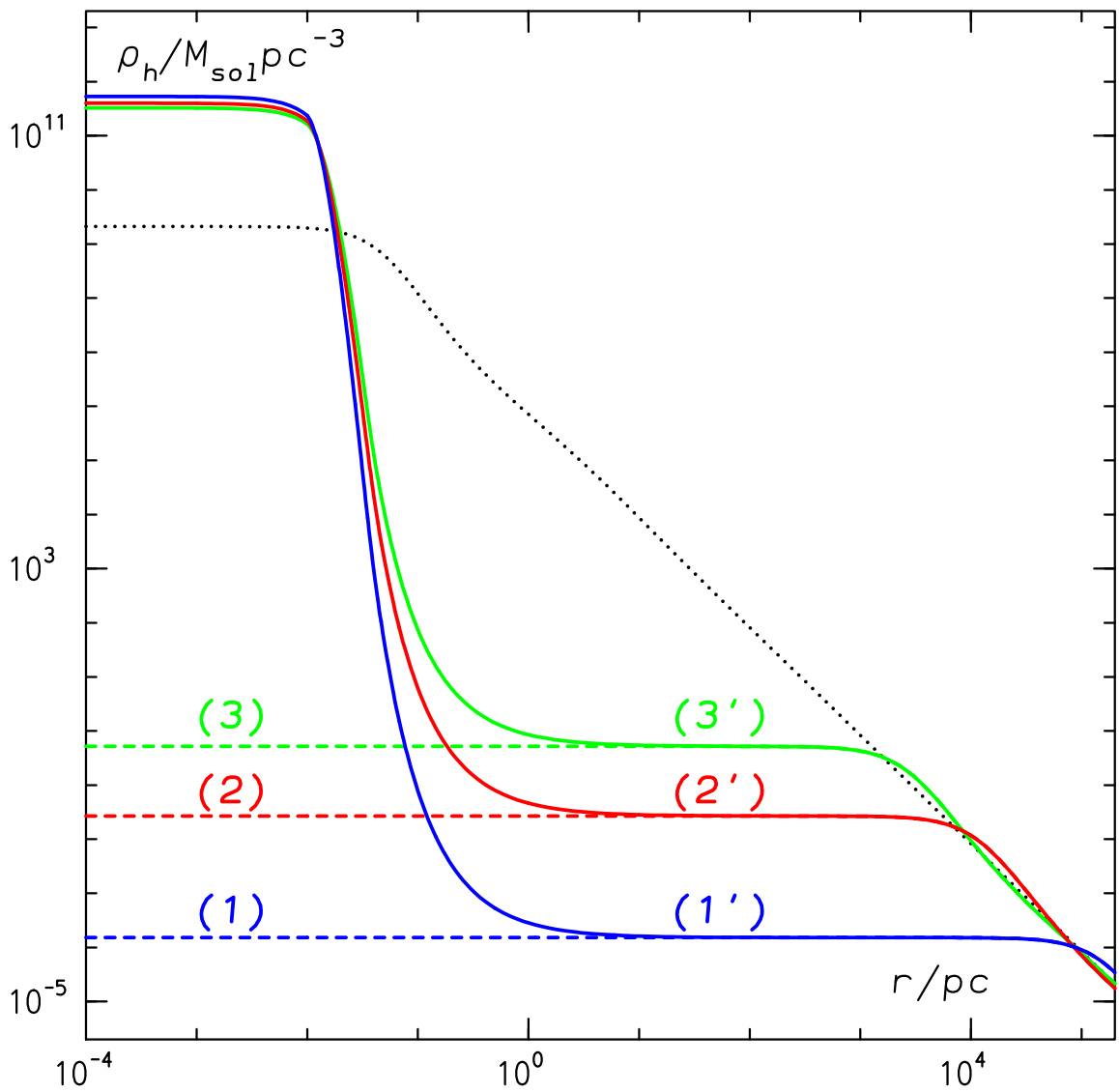
\begin{figure}
\caption{
The density profile of the halo
for a central degeneracy parameter $\eta_0=0$ (dotted line) and for
the six $\eta_0$-values discussed in the text.
Configurations with negative $\eta_0$
((1)-(3)) are depicted by the dashed
and those with positive $\eta_0$
((1')-(3')) by the solid line.
}
\label{fig6}
\end{figure}
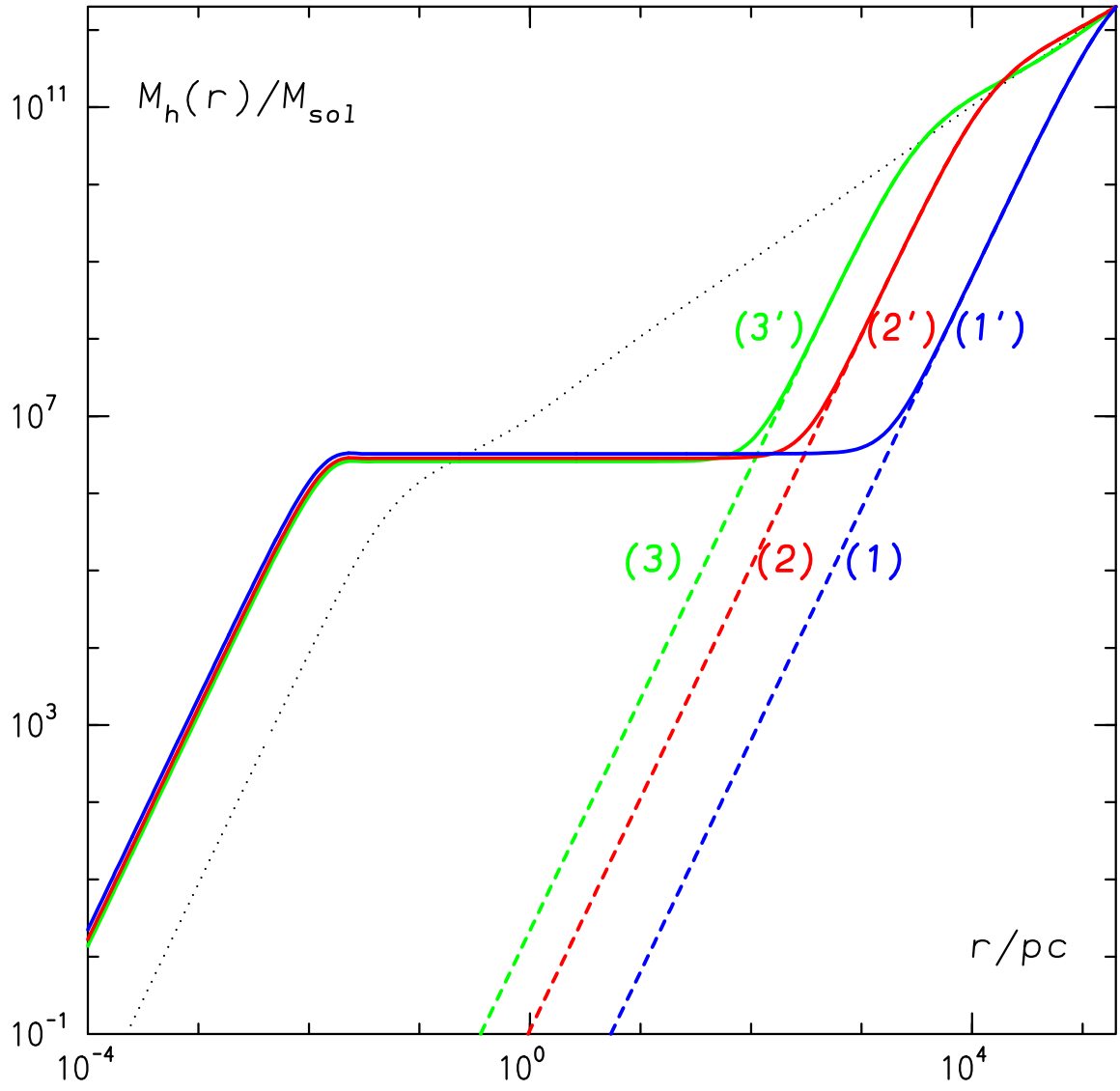
\begin{figure}
\caption{
Mass of the halo $M_{\rm h}(r)$ enclosed within a radius $r$
for various central degeneracy parameters
$\eta_0$ as in Fig. \protect\ref{fig6}.
}
\label{fig7}
\end{figure}
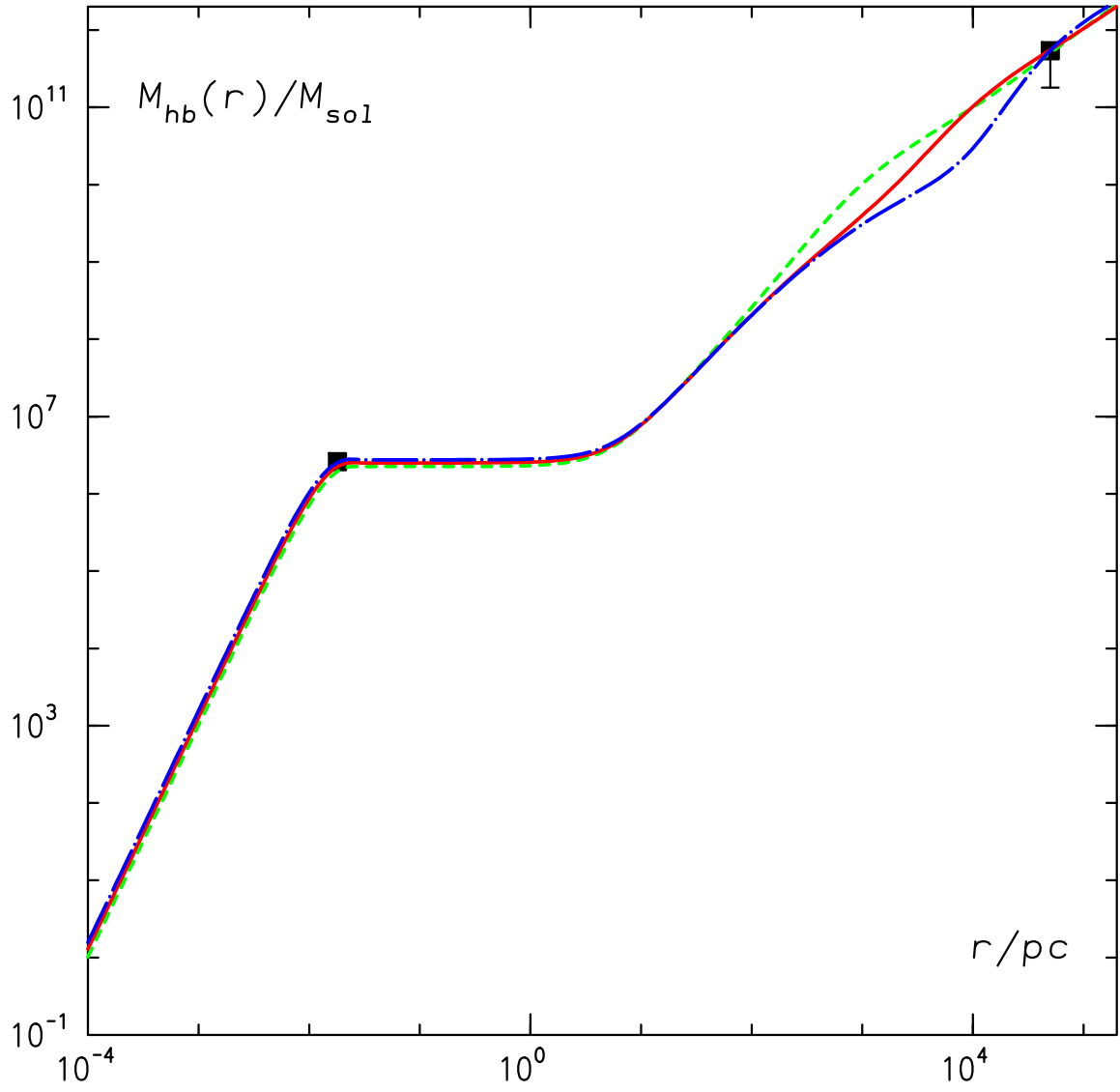
\begin{figure}
\caption{
Enclosed mass of halo plus bulge versus
radius for $\eta_0$ =
   24 (dashed),
  28 (solid),
and
    32  (dot-dashed line).
}
\label{fig8}
\end{figure}
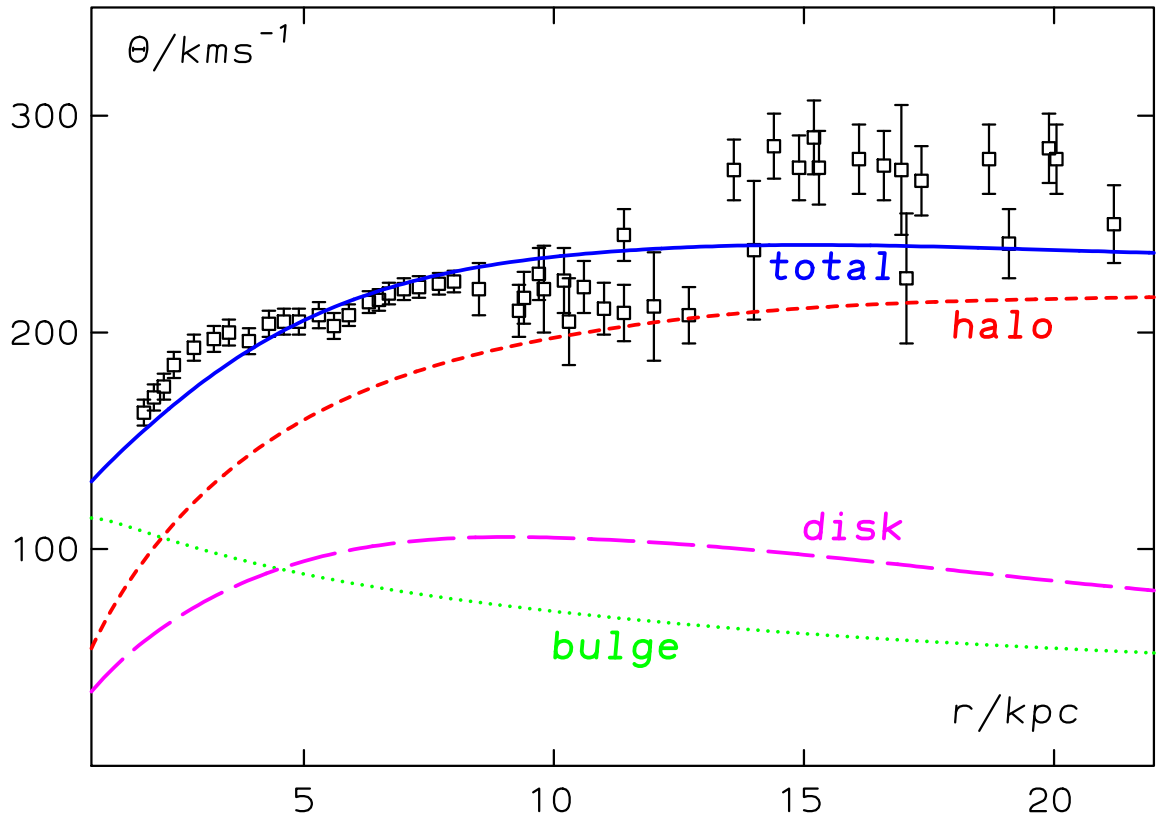
\begin{figure}
\caption{
Fit to the rotation curve of the Galaxy.
The data points are from \protect\cite{oll}
for $R_0=8.5$ kpc and $\Theta_0=220$ km/s.
}
\label{fig9}
\end{figure}
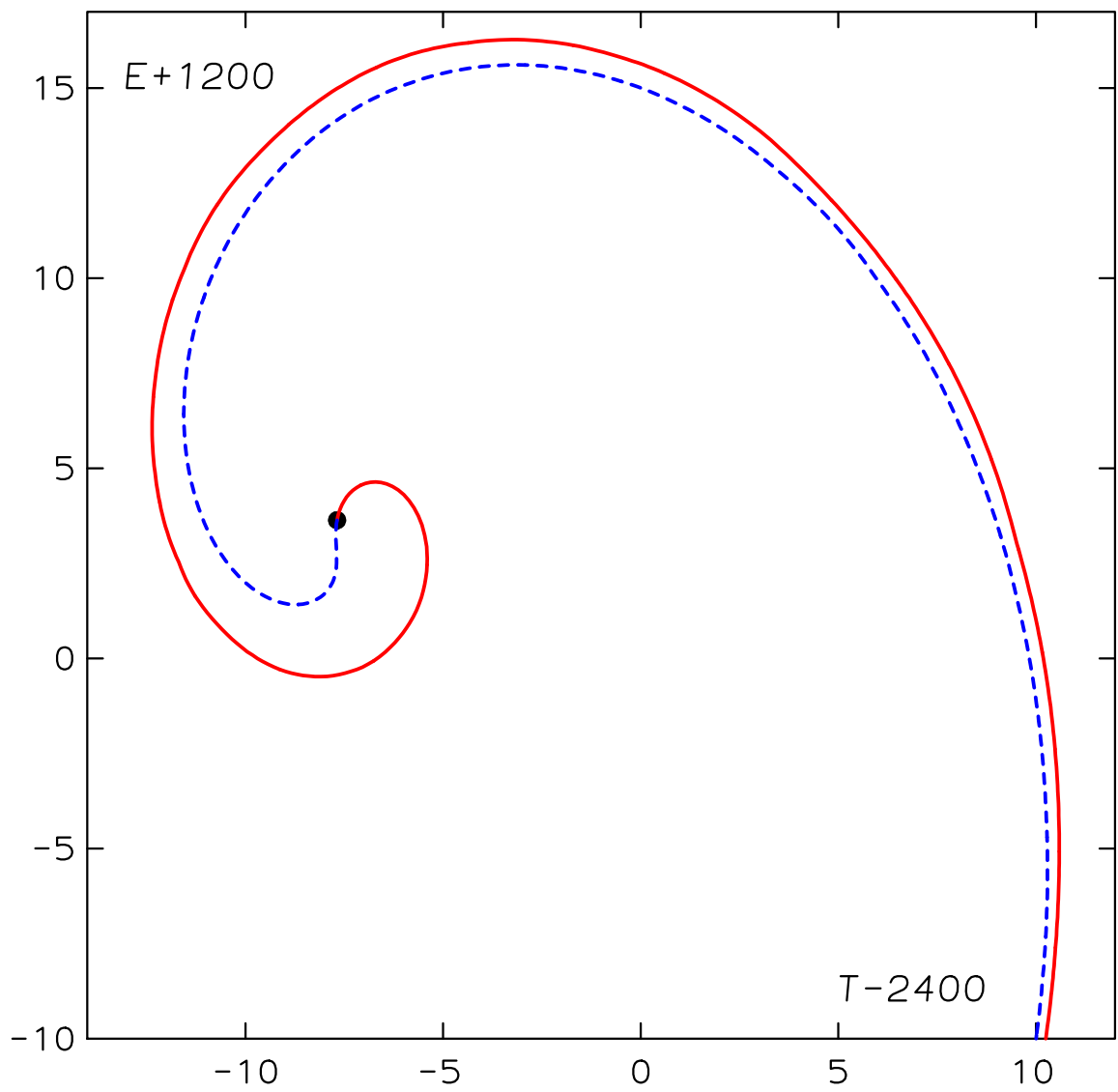
\begin{figure}
\caption{
Energy (shifted by $12\times 10^{-8}m$) versus
temperature (shifted by $-24\times 10^{-8}m$), both in
units of $10^{-10}m$, for fixed $N=2\times 10^{12} M_{\odot}/m$
}
\label{fig10}
\end{figure}
%
\end{document}